  \documentclass[12pt,preprint]{aastex}
\shorttitle{Modified equipartition calculation for supernova
remnants} \shortauthors{Arbutina et al.}

\begin{document}

\title{Modified equipartition calculation for supernova
remnants. Cases $\alpha =0.5$ and $\alpha =1$}

\author{B. Arbutina$^1$, D. Uro\v{s}evi\'{c}$^{1,2}$, M. M. Vu\v{c}eti\'{c}$^1$, M. Z.
Pavlovi\'{c}$^1$ and B. Vukoti\'{c}$^3$}

\affil{$^1$Department of Astronomy, Faculty of Mathematics,
University
of Belgrade, Studentski trg 16, 11000 Belgrade, Serbia;  {arbo@math.rs}\\
$^2$Isaac Newton Institute of Chile, Yugoslavia Branch\\
$^3$Astronomical Observatory, Volgina 7, 11060 Belgrade 38,
Serbia}

\begin{abstract}
The equipartition or minimum-energy calculation is a well-known
procedure for estimating magnetic field strength and total energy
in the magnetic field and cosmic ray particles by using only the
radio synchrotron emission. In one of our previous papers we have
offered a modified equipartition calculation for supernova
remnants (SNRs) with spectral indices $0.5<\alpha <1$. Here we
extend the analysis to SNRs with $\alpha =0.5$ and $\alpha =1$.
\end{abstract}


\section{Introduction}

The equipartition or minimum-energy calculation is a well-known
procedure for estimating magnetic field strength and total energy
in the magnetic field and cosmic ray particles from the radio
synchrotron emission  of a source. It is often used when no other
methods are available. Details of classical equipartition and
revised equipartition calculations for radio sources in general
are available in Pacholczyk (1970) and Beck \& Krause (2005),
respectively. Lacki \& Beck (2013) derived equipartition formula
for starburst galaxies.  In one of our previous papers (Arbutina
et al. 2012, hereafter Paper I) we have offered a modified
equipartition calculation for supernova remnants (SNRs) with
spectral indices $0.5<\alpha <1$. Spectral index is defined
through equation $S_\nu \propto \nu ^{-\alpha}$, where $S_\nu$ is
the flux density.
 Our approach was similar to that of
Beck \& Krause (2005). However, rather than introducing a
break in the power-law energy distribution, we assumed power-law spectra and
integrated over momentum to obtain energy densities of particles.
We further took into account different ion species
(not just protons and electrons), used flux density at a given
frequency, assumed isotropic distribution of the pitch angles
for the remnant as a whole, and incorporated the dependence on
shock velocity $v_s$ through injection energy $E_{\mathrm{inj}}
\sim m_p v_s ^2$.

For simplicity, in Paper I we have assumed that cosmic rays (CRs)
momentum/energy spectrum extends to infinity. This introduces only
small error in the final result because of the power-law
dependence of the momentum distribution function $f(p) = k
p^{-(\gamma +2)}$. However, if $\gamma$ is exactly 2, the integral
for cosmic rays energy density diverges and one has to set an
upper limit for particle energy. This is unfortunate, since
$\alpha =0.5$ ($\gamma = 2\alpha +1$) is considered to be a
typical spectral index for supernova remnants. Out of 40 SNRs for
which we can estimate magnetic field from equipartition
calculation, 22 have spectral index $\alpha=0.5$ (Pavlovi\'{c} et
al. 2013). This value comes directly from the theory of diffusive
shock acceleration (DSA) in the case of strong shocks with
compression ratio $r=4$; $\gamma = (r+2)/(r-1)$  (Bell 1978a).
Uro\v{s}evi\'{c} et al. (2012) tried to overcome the problem by
setting a fixed upper limit ($10^{15}$ eV for protons and
$10^{12}$ eV for electrons). In reality, however, CRs maximum
energy will also depend on magnetic field (and $v_s$) which  makes
the equipartition, i.e. minimum-energy calculation, more
complicated.

A similar situation arises in the case $\alpha =1$ -- integral for
cosmic rays energy density diverges unless a lower limit for
particle energy, i.e. $E_{\mathrm{inj}}$, is set. SNRs generally
have spectral indices lower than 1. Still, SN 1987A remnant, for
example, had $\alpha =1$ and higher between days $\sim$2000--3000
since explosion (Zanardo et al. 2010).
 In the next section we will
consider special cases corresponding to $\alpha =0.5$ and $\alpha
=1$.

\section{Analysis and Results}

All designations in this section are taken from Paper I.

\subsection{Case $\alpha =0.5$}

Let us start with energy density of a cosmic ray species with
$\gamma = 2$ ($\alpha = 0.5$)
\begin{eqnarray}
\epsilon  &=& \int _{p_{\mathrm{inj}}} ^{p_\infty} 4\pi k p^{-2}
(\sqrt{p^2c^2+m^2c^4}-mc^2)\mathrm{d}p
\\
&= & K \int _{x_{\mathrm{inj}}} ^{x_\infty} x^{-2} (\sqrt{x^2 +
1}-1)\mathrm{d}x , \ \ \ x=\frac{p}{mc},\ K= 4\pi kc , \nonumber
\\
&= & K\ I_2(x) | _{x_{\mathrm{inj}}} ^{x_\infty} = K\
\Big(\frac{1}{x}-\frac{\sqrt{x^2 +1}}{x}+ \mathrm{arcsinh}\
x\Big)\Big| _{x_{\mathrm{inj}}} ^{x_\infty}  , \nonumber
\end{eqnarray}
where $k$ and $K$ are the constants in the momentum and energy
distribution functions, respectively. When $x\rightarrow 0$,
$I_2(x) \approx x/2$ while for $x\rightarrow \infty$, $I_2(x)
\approx \ln x$. Total cosmic ray energy density $\epsilon
_{\mathrm{CR}} = \epsilon _{\mathrm{e}} + \epsilon
_{\mathrm{ion}}$ is then
\begin{eqnarray}
\epsilon _{\mathrm{CR}} &=&  K_e \Big( I_2\big( {\scriptstyle
\frac{p_\infty ^e}{m_e c} } \big) -  I_2\big( {\scriptstyle
\frac{p_\mathrm{inj} ^e}{m_e c} } \big) \Big) \nonumber \\
&+& \sum _i K_i \Big( I_2\big( {\scriptstyle \frac{p_\infty
^i}{m_i c} } \big)
-  I_2\big( {\scriptstyle \frac{p_\mathrm{inj} ^i}{m_i c} } \big) \Big) \nonumber \\
& \approx & K_e \bigg\{ \ln \big( \frac{E_\infty ^e}{m_e c^2} \big) - I_2 \big( {\scriptstyle \frac{\sqrt{E_{\mathrm{inj}}^2 +2m_e c^2 E_{\mathrm{inj}}}}{m_e c^2}} \big) + \frac{1}{\sum _i Z_i \nu _i}  \nonumber \\
&\cdot&   {\scriptstyle \sqrt{\frac{2m_p c^2
E_{\mathrm{inj}}}{E_{\mathrm{inj}}^2+ 2m_e c^2 E_{\mathrm{inj}}}}}
\Big(  \sum _i \sqrt{A_i} \nu _i \ln \big( \frac{E_\infty ^p}{m_p
c^2} \big) - \sqrt{\frac{E_{\mathrm{inj}}}{2m_p c^2}}\nonumber \\
&+&  \sum _i \sqrt{A_i} \ln \big( {\scriptstyle
\frac{Z_i}{A_i}}\big) \nu _i \Big) \bigg\}.
\end{eqnarray}
where we followed Bell's (1978a,b) DSA theory and his assumptions
concerning injection of particles into the acceleration process,
used equations (25) and (26) from Paper I and assumed
$E_{\mathrm{inj}}\ll m_p c^2$, $p_\infty \approx E_\infty /c$ for
all CRs species. $A_i$ and $Z_i$ are mass and charge numbers,
respectively, and $\nu _i$ represent ion abundances (for other
details see Paper I). Assuming Bohm diffusion and synchrotron
losses for electrons, for maximum electron energy we use $E_\infty
^e = \frac{3}{8} \frac{m_e^2 c^3 v_s}{\sqrt{\frac{2}{3}e^3B}}$
(Zirakashvili \& Aharonian 2007) and for ions $E_\infty ^i =
\frac{3}{8}\frac{v_s}{c}Z_i e B R$ (see Bell et al. 2013 and
references therein), where $R$ is SNR radius. Both formulae are in
cgs units. In reality, of course, we do not expect a sharp break
in the energy spectra, but some steepening, especially in the case
of electrons (Blasi 2010).

For the total energy we have $ E = \frac{4\pi}{3}R^3 f (\epsilon
_{\mathrm{CR}} + \epsilon _B)$, $\epsilon _B = \frac{1}{8\pi}
B^2$. Looking for the minimum energy with respect to $B$,
$\frac{\mathrm{d}E}{\mathrm{d}B} =0$ gives
\begin{eqnarray}
&&\frac{\mathrm{d} K_e}{\mathrm{d}B} \bigg[ \ln \big(
\frac{E_\infty ^e}{m_e c^2} \big) - I \big( {\scriptstyle
\frac{\sqrt{E_{\mathrm{inj}}^2 +2m_e c^2 E_{\mathrm{inj}}}}{m_e
c^2}} \big) +\frac{1}{3} + \frac{1}{\sum _i Z_i \nu _i} \nonumber
\\
&\cdot&   {\scriptstyle \sqrt{\frac{2m_p c^2
E_{\mathrm{inj}}}{E_{\mathrm{inj}}^2+ 2m_e c^2 E_{\mathrm{inj}}}}}
\Big(  \sum _i \sqrt{A_i} \nu _i \ln \big( \frac{E_\infty ^p}{m_p
c^2} \big) - \sqrt{\frac{E_{\mathrm{inj}}}{2m_p c^2}}\nonumber \\
&+&  \sum _i \sqrt{A_i} \ln \big( {\scriptstyle
\frac{Z_i}{A_i}}\big) \nu _i  -\frac{2}{3} \sum _i \sqrt{A_i} \nu
_i \Big) \bigg] + \frac{1}{4\pi} B  =0 .
\end{eqnarray}
where (by using equations (4), (5) and (6) from Paper I)
\begin{equation}
\frac{\mathrm{d} K_e}{\mathrm{d}B} = -\frac{3}{2} \frac{K_e}{B}= -
\frac{9}{4\pi} \frac{S_\nu}{f \theta ^3 d }\frac{1}{c_5}
\Big(\frac{\nu}{2c_1}\Big)^{1/2} \frac{\Gamma
(\frac{9}{4})}{\sqrt{\pi}\Gamma (\frac{7}{4})} B^{-5/2}.
\end{equation}
In order to find magnetic field equation (3) has to be solved
numerically. When one finds $B$, minimum energy can be obtained
from
\begin{equation}
E_{\mathrm{min}} = \Big( 1+ \frac{4}{3} \frac{\{\ldots
\}}{[\ldots]} \Big) E_B,\ \ \ E_B = \frac{4\pi}{3}R^3 f\ \epsilon
_B,
\end{equation}
where $\{ \ldots \}$ and $[\ldots ]$ are expressions in the
corresponding brackets in equations (2) and (3), respectively.

\subsection{Case $\alpha =1$}

In the situations when $\gamma = 3$ ($\alpha =1$), energy density
of a cosmic ray species is
\begin{eqnarray}
\epsilon  &=& \int _{p_{\mathrm{inj}}} ^{p_\infty} 4\pi k p^{-3}
(\sqrt{p^2c^2+m^2c^4}-mc^2)\mathrm{d}p
\\
 &=&  \frac{K}{mc^2}\ I_3(x) | _{x_{\mathrm{inj}}}
^{x_\infty}, \ \ \ x=\frac{p}{mc},\ K= 4\pi kc^2 , \nonumber \\
&=& \frac{K}{mc^2}\ \Big(\frac{1-\sqrt{x^2 +1}}{2 x^2}-\frac{1}{2}
\ln  \Big( \frac{1+\sqrt{x^2 +1}}{x}  \Big)\Big)\Big| _{x_{\mathrm{inj}}}
^{x_\infty}  . \nonumber
\end{eqnarray}
When $x\rightarrow 0$, $I_3(x) \approx \frac{1}{2} \ln x$ while
for $x\rightarrow \infty$, $I_3(x) \approx -\frac{1}{2x}$. Total
cosmic ray energy density is then
\begin{eqnarray}
\epsilon _{\mathrm{CR}} &=&  \frac{K_e}{m_e c^2} \Big( I_3\big( {\scriptstyle
\frac{p_\infty ^e}{m_e c} } \big) -  I_3\big( {\scriptstyle
\frac{p_\mathrm{inj} ^e}{m_e c} } \big) \Big) \nonumber \\
&+& \sum _i \frac{K_i}{m_i c^2}  \Big( I_3\big( {\scriptstyle \frac{p_\infty
^i}{m_i c} } \big)
-  I_3\big( {\scriptstyle \frac{p_\mathrm{inj} ^i}{m_i c} } \big) \Big) \nonumber \\
& \approx & \frac{K_e}{m_e c^2}  \bigg\{  -\frac{m_e c^2}{2 E_\infty ^e}  - I_3 \big( {\scriptstyle \frac{\sqrt{E_{\mathrm{inj}}^2 +2m_e c^2 E_{\mathrm{inj}}}}{m_e c^2}} \big) - \frac{1}{\sum _i Z_i \nu _i}  \nonumber \\
&\cdot&   {\scriptstyle {\frac{m_e c^2
E_{\mathrm{inj}}}{E_{\mathrm{inj}}^2+ 2m_e c^2 E_{\mathrm{inj}}}}}
\Big( \sum _i \frac{A_i}{Z_i}    \frac{m_p
c^2}{ E_\infty ^p} \nu _i \nonumber + \frac{1}{2} \ln \big( \frac{2 E_{\mathrm{inj}}}{m_p c^2} \big) \\
&-&  \frac{1}{2} \sum _i  \ln ( {A_i} ) \nu _i \Big) \bigg\},
\end{eqnarray}
where we have used the same assumptions as in derivation of equation (2).

Derivative of the total energy with respect to $B$ gives
\begin{eqnarray}
&&\frac{1}{m_e c^2}  \frac{\mathrm{d} K_e}{\mathrm{d}B} \bigg[ -\frac{3 m_e c^2}{8 E_\infty ^e}  - I_3 \big( {\scriptstyle \frac{\sqrt{E_{\mathrm{inj}}^2 +2m_e c^2 E_{\mathrm{inj}}}}{m_e c^2}} \big) - \frac{1}{\sum _i Z_i \nu _i}  \nonumber \\
&\cdot&   {\scriptstyle {\frac{m_e c^2
E_{\mathrm{inj}}}{E_{\mathrm{inj}}^2+ 2m_e c^2 E_{\mathrm{inj}}}}}
\Big( \frac{3}{2} \sum _i \frac{A_i}{Z_i}    \frac{m_p
c^2}{ E_\infty ^p} \nu _i \nonumber + \frac{1}{2} \ln \big( \frac{2 E_{\mathrm{inj}}}{m_p c^2} \big) \\
&-&  \frac{1}{2} \sum _i  \ln ( {A_i} ) \nu _i \Big) \bigg] + \frac{1}{4\pi} B  =0 .
\end{eqnarray}
where (see equations (4), (5) and (6) from Paper I)
\begin{equation}
\frac{\mathrm{d} K_e}{\mathrm{d}B} = -2 \frac{K_e}{B}= -
\frac{9}{8\pi} \frac{S_\nu}{f \theta ^3 d }\frac{1}{c_5}
\frac{\nu}{c_1}  B^{-3}.
\end{equation}
To find magnetic field more precisely, equation (8) has to be
solved numerically. Nevertheless, unlike the case $\alpha=0.5$,
the solution will only weakly depend on the upper limits for
energy, so the terms containing $E_\infty ^e$ and $E_\infty ^p$
could, in principle, be neglected. When one finds $B$, minimum
energy can be obtained from
\begin{equation}
E_{\mathrm{min}} = \Big( 1+  \frac{\{\ldots
\}}{[\ldots]} \Big) E_B,\ \ \ E_B = \frac{4\pi}{3}R^3 f\ \epsilon
_B,
\end{equation}
where now $\{ \ldots \}$ and $[\ldots ]$ are expressions in the
corresponding brackets in equations (7) and (8), respectively.

\section{Conclusions}

In Paper I we have offered a modified equipartition calculation
for SNRs with spectral indices $0.5<\alpha <1$. In this paper we
extend the analysis to SNRs with $\alpha =0.5$ and $\alpha =1$.
Spectral indices higher than $\alpha =1$ are rarely observed in
SNRs and are more typical for some extended radio galaxies and
active galactic nuclei (Kellermann \& Owen 1988). Lower spectral
indices $\alpha <0.5$ can be expected  in SNRs expanding in a
low-$\beta$ plasma  (i.e. dominant magnetic field, where $\beta$
is the ratio of thermal to magnetic pressures, Schlickeiser \& F\"
urst 1989). There are other possible explanations for SNRs with
$\alpha <0.5$ such as non-negligible thermal emission (Oni\' c
2013). However, the assumption of "equipartition" is likely to be
no longer valid in these cases. Case $\alpha =0.5$ is particularly
important since this value comes directly from test-particle DSA
theory in the case of strong shocks. Of course, non-linear DSA
theory does not give simple power-law spectra so we are always
talking about average spectral indices. In Table 1 we have
calculated magnetic field strengths for 22 SNRs by applying four
different methods. Flux densities at 1 GHz, angular sizes and
distances are taken from Green (2009) and Pavlovi\'{c} et al.
(2013). Our values for SNRs with $\alpha=0.5$ are approximately
40\% higher than those derived by using classical approach of
Pacholczyk (1970) and similar to those obtained by applying Beck
\& Krause (2005) revised equipartition formula and
Uro\v{s}evi\'{c} et al. (2012) approximation. Nevertheless, the
derivation presented in this paper is more accurate than the
latter two -- in particular, the upper limit for particle energy
i.e. momentum depends on the magnetic field itself and the upper
and lower limits depend on shock velocity, and by varying this
last parameter one can obtain different magnetic field estimates.

The Web application for calculation of the magnetic field strength
of SNRs is available at {\it
http://poincare.matf.bg.ac.rs/\~{}arbo/eqp/}.

\acknowledgments

The authors acknowledge financial support of the Ministry of
Education, Science and Technological Development of the Republic
of Serbia through the projects 176004 'Stellar physics', 176005
'Emission nebulae: structure and evolution' and 176021 'Visible
and invisible matter in nearby galaxies: theory and observations'.

\begin{deluxetable}{@{\extracolsep{-3mm}}llcccc@{}}
\tabletypesize{\small} \tablecolumns{6} \tablecaption{Calculated
magnetic field strengths for SNRs with $\alpha=0.5$} \vskip -2mm
\tablehead{
\colhead{} & \colhead{} & \colhead{Pacholczyk (1970)} & \colhead{Beck\&Krause (2005)} & \colhead{Uro\v{s}evi\'{c} et al. (2012)} & \colhead{This paper$^{*}$} \\
\multicolumn{1}{l}{Catalog name} & \multicolumn{1}{l}{Other name}
& \colhead{($\mu$Ga)} & \colhead{($\mu$Ga)} & \colhead{($\mu$Ga)}
& \colhead{($\mu$Ga)} }
\startdata
      G21.8-0.6     &   Kes 69  &   83.5    &    116.4      &    113.6    &      116.9     \\
        G23.3-0.3   &   W41  &     68.9     &    96.0       &    93.8    &      96.3     \\
      G33.6+0.1       &   Kes 79  &   100.3   &   139.7     &     136.4   &       139.7   \\
         G46.8-0.3  &   HC30  &     60.8    &    84.7       &     82.7   &    84.8      \\
       G54.4-0.3    & HC40    &   41.0      &   57.1        &    55.8    &      56.2    \\
         G84.2-0.8   &     &     56.7   &    79.0       &   77.1     &        78.3   \\
         G96.0+2.0  &     &    15.2     &     21.2      &    20.7    &    20.5      \\
         G108.2-0.6  &     &    19.8    &     27.6      &    26.9    &      27.3   \\
        G109.1-1.0   &   CTB 109  &     51.9    &    72.3       &    70.6    &     71.7      \\
       G114.3+0.3    &     &     24.5   &    34.1       &    33.3    &     32.8     \\
        G116.5+1.1   &     &   23.2     &      32.4     &   31.6     &     31.7     \\
         G156.2+5.7  &     &    14.7    &     20.4      &    19.9    &       19.8   \\
        G205.5+0.5   &   Monoceros Nebula  &     20.7   &   28.8        &  28.1      &      28.7    \\
         G260.4-3.4  &    Puppis A &    54.0    &    75.2       &    73.4    &     75.1     \\
         G292.2-0.5  &     &    42.9    &    59.7       &    58.3    &      59.6    \\
         G296.5+10.0  &    PKS 1209-51/52 &    30.9     &     43.1      &    42.0    &     42.8     \\
         G309.8+0.0  &     &    57.8    &    80.5       &    78.6    &    79.7      \\
         G332.4-0.4  &  RCW 103   &     135.6   &      188.9    &    184.4    &      186.8     \\
       G337.8-0.1    &  Kes 41   &    108.4     &    151.0      &   147.4     &    151.8      \\
       G344.7-0.1  &     &    55.5    &     77.3      &    75.5    &    76.2      \\
         G346.6-0.2  &     &    79.9    &     111.3     &   108.7     &     111.4     \\
           G349.7+0.2     &        &    274.7   &     382.7     &   373.7     &     375.2     \\
\enddata
\vskip 0mm \tablenotetext{\ }{$^{*}$For $\upsilon_{s} \sim $1000
km/s.}

\end{deluxetable}


\begin{thebibliography}{}


\bibitem[\protect\citeauthoryear{Arbutina et al.}{2012}]{Arbutina} Arbutina B., Uro\v{s}evi\'{c} D., Andjeli\'{c} M.M.,  Pavlovi\'{c} M.Z., Vukoti\'{c} B., 2012, \apj, 746, 79
(Paper I)

\bibitem[\protect\citeauthoryear{Beck \& Krause}{2005}]{BeckKrause} Beck R., Krause M., 2005, AN, 326, 414

\bibitem[\protect\citeauthoryear{Bell}{1978a}]{Bella} Bell A. R., 1978a, \mnras, 182, 147


\bibitem[\protect\citeauthoryear{Bell}{1978b}]{Bellb} Bell A. R., 1978b, \mnras, 182, 443

\bibitem[\protect\citeauthoryear{Bell et al.}{2013}]{Bell13} Bell A.R., Schure K.M., Reville B., Giacinti G.,  2013, \mnras, 431,
415

\bibitem[\protect\citeauthoryear{Blasi}{2010}]{Blasi} Blasi P., 2010, \mnras, 402,
2807

\bibitem[\protect\citeauthoryear{Green}{2009}]{Green} Green D. A., 2009, "A Catalogue of Galactic Supernova Remnants (2009 March version)", Astrophysics Group, Cavendish Laboratory, Cambridge, United Kingdom (available at http://www.mrao.cam.ac.uk/surveys/snrs/)

\bibitem[\protect\citeauthoryear{Kellermann}{1988}]{Kellermann} Kellermann I. K., Owen F. N., 1988, in "Galactic and Extragalactic Radio Astronomy (2nd edition)" eds. G. L. Verschuur and K. I. Kellermann, Berlin and New York: Springer-Verlag, p. 563

\bibitem[\protect\citeauthoryear{Lacki}{2013}]{Lacki} Lacki B. C. \& Beck R., 2013, \mnras, 430, 3171

\bibitem[\protect\citeauthoryear{Onic}{1989}]{Onic} Oni\' c  D., 2013, Ap\&SS, 346, 3


\bibitem[\protect\citeauthoryear{Pacholczyk}{1970}]{Pacholczyk}
Pacholczyk A. G., 1970, Radio Astrophysics, San Francisco: Freeman
and Co.

\bibitem[\protect\citeauthoryear{Pavlovic et al.}{2013}]{Pavlovic} Pavlovi\'{c} M. Z., Uro\v{s}evi\'{c} D., Vukoti\'{c} B., Arbutina B., G\"{o}ker \"{U}. D.,
2013, \apjs, 204, 4

\bibitem[\protect\citeauthoryear{Schlickeiser}{1989}]{Schlickeiser} Schlickeiser R., F\" urst E., 1989, A\&A 219, 192

\bibitem[\protect\citeauthoryear{Urosevic et al.}{2012}]{Urosevic} Uro\v{s}evi\'{c} D., Pavlovi\' c  M.Z., Arbutina B., Dobard\v{z}i\'{c} A., 2012,
XXVIIth IAU General Assembly, SpS4 "New era for studying
interstellar and intergalactic magnetic fields", August 20-31,
2012, Beijing, China

\bibitem[\protect\citeauthoryear{Zanardo et al.}{2010}]{Zanardo} Zanardo G. et al., 2010, \apj, 710,
1515


\bibitem[\protect\citeauthoryear{Zirakashvili \& Aharonian}{2007}]{ZA} Zirakashvili V.N., Aharonian F., 2007, A\&A, 465,
695


\end{thebibliography}
\end{document}